%% file: main.tex
\documentclass{article} 
\usepackage{spconf}
\usepackage{amsmath,amssymb}
\usepackage{bm}
\usepackage{mathtools}
\usepackage{steinmetz}
\usepackage{graphicx,epsfig}
\usepackage{setspace} 
\usepackage{multirow}
\usepackage{mwe}
\usepackage{algpseudocode}
\usepackage[ruled,linesnumbered]{algorithm2e}
\usepackage{verbatim} 
\usepackage{subfigure}
\usepackage{cite}

\SetKwInput{KwInput}{Input}               
\SetKwInput{KwOutput}{Output}
\makeatletter
\def\endthebibliography{%
	\def\@noitemerr{\@latex@warning{Empty `thebibliography' environment}}%
	\endlist
}
\makeatother

\title{Supervised Learning based Sparse Channel Estimation for RIS aided Communications}
\name{Dilin~Dampahalage,~K. B. Shashika Manosha,~Nandana~Rajatheva,~and~Matti~Latva-aho\thanks{This work was supported by the Academy of Finland 6Genesis Flagship (grant 318927).}}
\address{Centre for Wireless Communications, Univeristy of Oulu, Finland}
\begin{document}
\maketitle

\begin{abstract}
An reconfigurable intelligent surface (RIS) can be used to establish line-of-sight (LoS) communication when the direct path is compromised, which is a common occurrence in a millimeter wave (mmWave) network. In this paper, we focus on the uplink channel estimation of a such network. We formulate this as a sparse signal recovery problem, by discretizing the angle of arrivals (AoAs) at the base station (BS). On-grid and off-grid AoAs are considered separately. In the on-grid case, we propose an algorithm to estimate the direct and RIS channels. Neural networks trained based on supervised learning is used to estimate the residual angles in the off-grid case, and the AoAs in both cases. Numerical results show the performance gains of the proposed algorithms in both cases.
\end{abstract}

\section{Introduction} \label{introduction}
\input{introduction.tex}

\section{System Model} \label{system}
\input{system.tex}

\section{Problem Formulation} \label{problem}
\input{problem.tex}

\section{Algorithm Development} \label{algorithm}
\input{algorithm.tex}

\section{Results} \label{results}
\input{results.tex}

\section{Conclusion} \label{conclusion}
\input{conclusion.tex}

\bibliographystyle{IEEEbib} 
\bibliography{biblography.bib}

\end{document}

%% file: introduction.tex
Reconfigurable intelligent surfaces (RISs) have resulted a new wave of research towards the reconfigurability of the wireless propagation environment, with software controlled scattering of the incoming waves. The phase shifts of the individual elements can be configured smartly to achieve favourable conditions, for a given application. The passive operation and lower cost of these devices attracts more and more applications \cite{dilin_ris, capcity, power_transfer}.

The smart reconfigurability of an RIS depends on the availability of channel state information (CSI). An RIS consists of a large number of passive reconfigurable elements in general, which results in a large number of propagation paths in addition to the direct path. Hence, the channel estimation for RIS aided systems is challenging. Grouping of reconfigurable elements \cite{irs_ofdm}, and relying on angle of arrival (AoA) of the line-of-sight (LoS) base station (BS)-RIS channel and the angle of departure (AoD) of the LoS RIS-User \cite{irs_gmd} for passive beamforming has been used to overcome some of these challenges.

Channel estimation for RISs requires the design of activation patterns for the RIS elements. The impact of RIS activation pattern on the channel estimation performance is investigated in \cite{optimal_estimation}, where an optimal codebook is proposed based on a minimum variance unbiased estimator. The angular domain sparsity in millimetre wave (mmWave) channels can be exploited to reduce the dimensionality of the parameter space, which results in a reduced pilot overhead.   A sparse representation of the concatenated BS-IRS-user (cascaded) channel is derived in \cite{Wang2020}, which convert channel estimation into a sparse signal recovery problem. The  double-structured  sparsity  of the  angular  cascaded  channels is leveraged in \cite{Wei2021} to propose a double-structured orthogonal matching pursuit (DS-OMP)  based  cascaded channel estimation  scheme.

Compared to optimization based approaches in traditional communication algorithms, neural networks (NNs) provides the ability to learn complex patterns from data itself. This is beneficial in situations where the development of an algorithm is difficult due to the complexity of the problem. Supervised learning is applied for training when a labelled data set is available, where the outputs corresponds to the inputs are tabulated. Some applications of NNs related to channel estimation can be found in \cite{n1,n2,n3,n4}.

A major use case of RIS is to establish LoS through the reflected path when the user does not have LoS with the BS, but with the RIS. In this case, the joint channel from user to BS contains a specular component dominated by the reflected LoS link through RIS and, a scattering component from the direct Non-LoS (NLoS) link. A novel channel model for a similar scenario is proposed in \cite{novel_ris}, where a compact expression for the channel distribution is derived. 

In our work we consider the uplink of a mmWave network, where an RIS is used to assist the communication. We focus on utilizing an RIS in a scenario where the user lacks LoS with the BS. Based on this model, we develop a compact representation for the RIS channel. An angular domain sparse channel model is considered by discretizing the AoAs. The channel estimation problem is formulated step by step for the case where AoAs lie exactly on the discrete grid (on-grid), and the case where AoAs can take any discrete value deviating from discrete grid (off-grid). A sparse estimation method is proposed for the on-grid case, based on OMP, and a NN based approach is used for comparison.  In the off-grid case, a two-step procedure is used to perform channel estimation, where first, on-grid AoAs are estimated using a NN and then off-grid AoAs are calculated by predicting the residuals using another NN.



%% file: system.tex
\begin{figure}[ht]
	\centerline{\includegraphics[width=0.5\columnwidth]{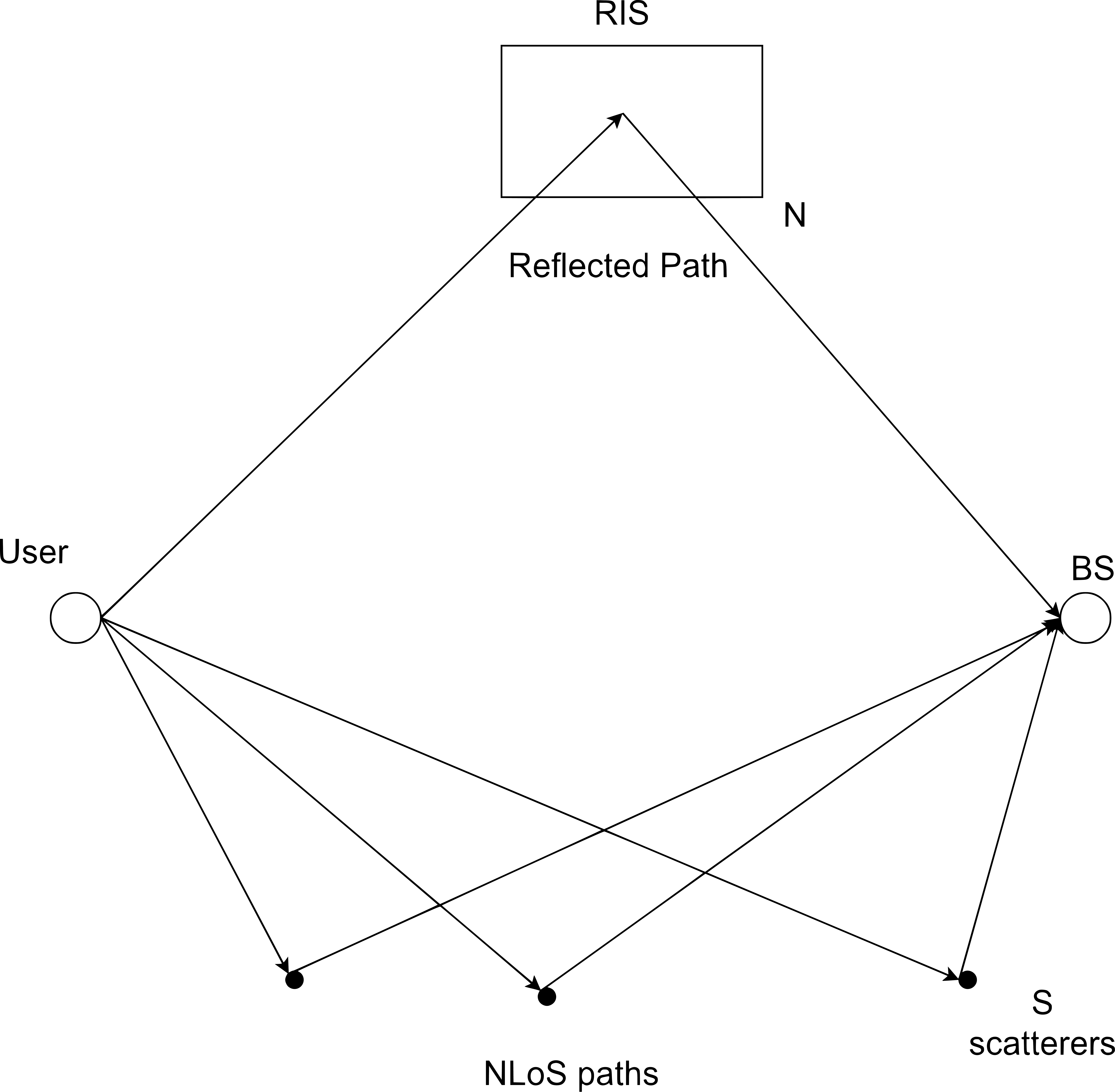}}
	\caption{Illustration of the system model.}
	\label{fig:system_model}
\end{figure}

We consider the uplink channel estimation of a mmWave network consisting of a single antenna user and a BS with $M$ antennas. An RIS with $N$ reflecting elements is used to assist the communication, while the reflecting elements are arranged in $G$ groups to reduce the training overhead \cite{dilin_ris}. As illustrated in Fig. \ref{fig:system_model}, the channel consists of a direct link which is non-line-of sight (NLoS), and a link through RIS which consist of line-of-sight (LoS) components. Let $\bm{H}_r \in \mathbb{C}^{M \times N}$ be the channel between RIS and BS,  $\bm{h}_v \in \mathbb{C}^{N}$ be the channel between user and RIS, and  $\bm{h}_d \in \mathbb{C}^{M}$ be the direct channel between user and BS. Let $\bm{\Theta}=\text{diag}(\bm{v})$ denote the reflection matrix of the RIS, where $\bm{v}=[e^{j\alpha_1},e^{j\alpha_2},...,e^{j\alpha_N}]^T$ with $\alpha_{n}$ being the phase shift of the $n$th reflecting element.


Since, user-RIS and RIS-BS links are LoS, we can express them as follows, in terms of the array responses of the RIS and BS,
\begin{equation}
\bm{h}_v = c_v \bm{a}_{RIS}(\theta_R^{RIS}),
\label{eqn:h_v}
\end{equation}
and
\begin{equation}
\bm{H}_r = c_r \bm{a}_{BS}(\theta_R^{BS}) \left[\bm{a}_{RIS}(\theta_T^{RIS})\right]^H,
\label{eqn:H_r}
\end{equation}
where $\bm{a}_{RIS}()$, $\bm{a}_{BS}()$ are the array responses of BS and RIS, and 
$\theta_R^{RIS}$, $\theta_R^{BS}$ and $\theta_T^{RIS}$ are the AoA at RIS, AoA at the BS and AoD at the RIS respectively. Here, $c_r$ and $c_v$ are the complex path gains of the links. A planar geometry is assumed in the derivations. Now, the effective channel through RIS can be expressed as,
\begin{equation}
\begin{split}
\bm{g} &= \bm{H}_r \bm{\Theta} \bm{h}_v\\
&=c_r \bm{a}_{BS}(\theta_R^{BS}) \left[\bm{a}_{RIS}(\theta_T^{RIS})\right]^H c_v \bm{\Theta} \bm{a}_{RIS}(\theta_R^{RIS})\\ 
&= \bm{a}_{BS}(\theta_R^{BS}) \left(\sum_{n=1}^{N}\gamma_{n}^* e^{j\alpha_n}\right)\\
&= \bm{a}_{BS}(\theta_R^{BS}) \bm{\gamma}^H\bm{v},
\end{split}
\label{eqn:g}
\end{equation}
where $\bm{\gamma}=[\gamma_1,\gamma_2,...,\gamma_N]^T$ with $\gamma_n = c_r c_v \left[a_{RIS,n}(\theta_T^{RIS})\right]$ $a_{RIS,n}(\theta_R^{RIS})^*$.

%% file: problem.tex
\subsection{On-Grid AoAs}
We assume a mmWave scattering channel model for the direct channel. The time is divided into frames, and in each frame, the user transmits several pilot symbols for channel estimation. The pilot symbols are generated by changing the phase shifts of the RIS by going through the entries of a predefined codebook \cite{optimal_estimation}. The effective channel $\bm{h}_{t,i}\in \mathbb{C}^{M}$ between the user and BS at the $i$th symbol of the $t$th frame is,
\begin{equation}
	\begin{split}
		\bm{h}_{t,i} &= \bm{g}_{t,i} + \sum_{q=1}^{S}\beta_{t,q} \bm{a}_R(\theta_{t,q})\\
		&= \bm{a}_{R}(\theta_{t,0}) \bm{\gamma}_t^H\bm{v}_{t,i} + \sum_{q=1}^{S}\beta_{t,q} \bm{a}_R(\theta_{t,q}),
	\end{split}
	\label{eqn:h_i}
\end{equation}
where $\bm{a}_R(\theta_{t,q})\in \mathbb{C}^{M}$ is the steering vector at the BS with $\theta_{t,q}$ being the AoA at the BS for the $q$th path, and $\beta_{t,q}$ is the complex path gain. Here, $\bm{g}_{t,i}$ is channel through RIS, $\theta_{t,0}$ corresponds to the AoA for the RIS-BS path and  $\bm{g}_{t,i}=\bm{\gamma}_t^H\bm{v}_{t,i}$, where $\bm{\gamma}_t$ is the complex path gain vector through RIS reflecting elements defined above. 

Let us consider an angular domain channel representation, where the angle of arrival is assumed to be from a grid of $K$ AoAs.  A dictionary $\bm{A}_R\in \mathbb{C}^{M \times (K+1)}$ is formed by the steering vectors of the grid points $\tilde{\theta}_k, \text{ for } k=0,1,\cdots,K$,
\begin{equation}
	\bm{A}_R=
	\begin{bmatrix}
		\bm{a}_R(\tilde{\theta}_0) & \bm{a}_R(\tilde{\theta}_1) & \cdots & \bm{a}_R(\tilde{\theta}_K)
	\end{bmatrix},
	\label{eqn:Ar}
\end{equation}
where $\tilde{\theta}_0$ corresponds to the AoA for RIS-BS link. Now we can write the channel vector as follows,
\begin{equation}
	\bm{h}_{t,i} = \bm{A}_R\left(\bm{z}'\bm{\gamma}_t^H\bm{v}_{t,i} + \bar{\bm{z}}_t\right),
	\label{eqn:h_i_ang}
\end{equation}
where $\bar{\bm{z}}_t\in\mathbb{C}^{K}$ is the sparse complex channel gain vector for the NLoS path, and $\bm{z}'$ is an all zeros vector except the first entry. The received signal at the $i$th time slot of the $t$th frame is,
\begin{equation}
	\bm{y}_{t,i} =  \bm{A}_R\left(\bm{z}'\bm{\gamma}_t^H\bm{v}_{t,i} + \bar{\bm{z}}_t\right)x_i + \bm{n}_{t,i},
	\label{eqn:y_i}
\end{equation}
where $\bm{n}_{t,i}=[n_{t,i,1},n_{t,i,2},\cdots,n_{t,i,M}]$ is the additive white Gaussian noise (AWGN) added at the BS antenna, with $n_{t,i,j} \sim \mathcal{CN}(0, N_0)$. For simplicity assume $x_i=1$. We observe the received signals while changing the phase shifts at the RIS for $L$ time slots. Let us also assume the channel parameters are not changing during this period. However, the complex channel gain through RIS changes depending on the phase shift configuration. Let $\bm{Q}_t=\bm{z}'\bm{\gamma}_t^H$, and we can see that $\bm{Q}\in\mathbb{C}^{K \times N}$ is a sparse matrix. Now we can write the received signal as,
\begin{equation}
	\bm{y}_{t,i} = \bm{A}_R(\bm{Q}_t\bm{v}_{t,i}+\bar{\bm{z}}_t) + \bm{n}_{t,i}.
	\label{eqn:y_i_2}
\end{equation}

We can see that $\bm{A}_R\bm{Q}_t$ and $\bm{A}_R\bar{\bm{z}}_t$  corresponds to $\bm{\Phi}_t$ and $\bm{h}_{d,t}$. For convenience we drop the frame index. Let $\bm{Y}=\begin{bmatrix}
	\bm{y}_1 & \bm{y}_2 & \cdots & \bm{y}_L
\end{bmatrix}$, $\bm{V}=\begin{bmatrix}
	\bm{v}_1 & \bm{v}_2 & \cdots & \bm{v}_{L}
\end{bmatrix}$ and $\bm{N}=\begin{bmatrix}
	\bm{n}_1 & \bm{n}_2 & \cdots & \bm{n}_L
\end{bmatrix}$.
Therefore, we can write the following matrix equation,
\begin{equation}
	\bm{Y} = \bm{A}_R\left(\bm{Q}\bm{V} + \bar{\bm{z}}\bm{\mathbf{1}}^T\right) + \bm{N}
	\label{eqn:Y_ongrid}
\end{equation}
We can express this in a compact notation by defining $\tilde{\bm{Q}}=\begin{bmatrix}
	\bar{\bm{z}} & \bm{Q}
\end{bmatrix}$ and $\tilde{\bm{V}}=\begin{bmatrix}
	\bar{\bm{\mathbf{1}}} & \bm{V}
\end{bmatrix}$. Now the Equation (\ref{eqn:Y_ongrid}) can be written as, 
\begin{equation}
	\bm{Y} = \bm{A}_R\tilde{\bm{Q}}\tilde{\bm{V}} + \bm{N}.
	\label{eqn:Y_ongrid2}
\end{equation}

\subsection{Off-Grid AoAs}

Although we have assumed on-grid AoAs, this is not the case in general. Therefore, in this section we modify the problem by relaxing on-grid AoA constraints. We introduce the residual AoA vector $\bm{\delta}=\begin{bmatrix}
	\delta_0 & \delta_1 & \cdots & \delta_K
\end{bmatrix}$ such that $\delta_k = \begin{cases}
\theta_q-\tilde{\theta}_{k_q}, & q=k_q \text{, for } q=0,1,\cdots,S \text{, and }k\neq0\\
0, & \text{otherwise},
\end{cases}$\\
where $k_q$ is the index of the grid point closest to the AoA of the $q$th path, and $\theta_q$, $\tilde{\theta}_{k_q}$ being the true AoA of the $q$th path and AoA value of the $k_q$th grid point respectively.
Now the Equation (\ref{eqn:Y_ongrid2}) can be modified to include the new BS array response dictionary that depends on $\bm{\delta}$ as $\bm{A}_R(\bm{\delta})$, which is obtained by correcting the grid points with residual values. We can express the received signal by the following matrix equation,
 \begin{equation}
 \bm{Y} = \bm{A}_R(\bm{\delta})\tilde{\bm{Q}}\tilde{\bm{V}} + \bm{N}.
 \label{eqn:Y_offgrid}
 \end{equation}

%% file: algorithm.tex
\subsection{Algorithm Development for On-Grid AoAs}
In this section we develop an algorithm to find the sparse solution of (\ref{eqn:Y_ongrid2}). In order to solve this problem we can use the known information regarding the sparsity of $\bm{Q}$ and $\bar{\bm{z}}$. We know that RIS act as single scarring element in our model, also it has a fixed location which can be perfectly known. On the other hand,  $\bar{\bm{z}}$ should be $S$-\textit{sparse} corresponding to the remaining scattering paths. 

Let us consider the expansion of the matrix product $\tilde{\bm{Q}}\tilde{\bm{V}}=\sum_{j=0}^N\tilde{\bm{q}}_j\tilde{\bm{v}}_j^T$, where $\tilde{\bm{q}}_j$ is the $j$th row of $\tilde{\bm{Q}}$ and $\tilde{\bm{v}}_j$ is the $j$th column of $\tilde{\bm{V}}$. Assume we have chosen $\tilde{\bm{V}}$ such that $\tilde{\bm{v}}_j^T\tilde{\bm{v}}_0^*=0\text{, for all }j\neq0$ (which is satisfied by the codebook in \cite{optimal_estimation}). Multiplying equation (\ref{eqn:Y_ongrid2}) by $\tilde{\bm{v}}_0^*$, we can eliminate all the other terms except $\tilde{\bm{q}}_0\text{ }(=\bar{\bm{z}})$,
\begin{equation}
	\begin{split}
		\bm{Y}\tilde{\bm{v}}_0^*&= \bm{A}_r\tilde{\bm{q}}_0\tilde{\bm{v}}_0^T\tilde{\bm{v}}_0^* + \bm{N}\tilde{\bm{v}}_0^*\\
		\bm{y}'&=\bm{A}_r\bar{\bm{z}} + \bm{n}',
	\end{split}
 \label{eqn:y'}
\end{equation}
where $\bm{y}'=\frac{\bm{Y}\tilde{\bm{v}}_0^*}{\tilde{\bm{v}}_0^T\tilde{\bm{v}}_0^*}$ and $\bm{n}'=\frac{\bm{N}\tilde{\bm{v}}_0^*}{\tilde{\bm{v}}_0^T\tilde{\bm{v}}_0^*}$. This corresponds to a sparse signal recovery problem and there exist many classical algorithms such as OMP, \cite{sparse_recovery} which can be used to estimate $\bar{\bm{z}}$.

After finding $\bar{\bm{z}}$, we can focus on the estimation of RIS channel by essentially removing the effect of the direct path from the received signal. We can define the signal due to RIS path as follows,
\begin{equation}
	\bm{Y}_\text{RIS} = \bm{Y} - \bm{A}_r\bar{\bm{z}}\tilde{\bm{v}}_0^T.
	\label{eqn:Y_ris}
\end{equation}
Since the AoA of the RIS-BS path is known, we can find $\bm{\gamma}$ by minimizing, 
\begin{equation}
	\|\bm{Y}_\text{RIS}-\bm{a}_R(\tilde{\theta}_0)\bm{\gamma}^H\bm{V}\|_2^2.
	\label{eqn:find_gamma}
\end{equation}
The channel estimation procedure is expressed in Algorithm \ref{alg:1}.
\begin{algorithm}
	\SetKwInOut{Input}{input}
	\SetKwInOut{Output}{output}
	\Indm
	\Input{Received signal $\bm{Y}$, AoA grid $\bm{A}_r$ and augmented $\tilde{\bm{v}}_0$ and RIS phase shifts $\tilde{\bm{V}}$.}
	\Output{Estimated $\bm{\Phi}$ and $\bm{h}_{d}$}
	\Indp
	\BlankLine
	Find $\bar{\bm{z}}$ by solving (\ref{eqn:y'}) with OMP\\
	Calculate $\bm{Y}_\text{RIS}$ by equation (\ref{eqn:Y_ris})\\
	Calculate $\bm{\gamma}$ by minimizing (\ref{eqn:find_gamma})\\
	$\bm{\Phi} = \bm{a}_R(\tilde{\theta}_0)\bm{\gamma}^H$\\
	$\bm{h}_d = \bm{A}_R\tilde{\bm{z}}$\\
	\caption{On-grid channel estimation algorithm}
	\label{alg:1}
\end{algorithm}

\subsection{Supervised Learning based Channel Estimation}
 
\begin{figure}[ht]
 	\centerline{\includegraphics[width=\columnwidth]{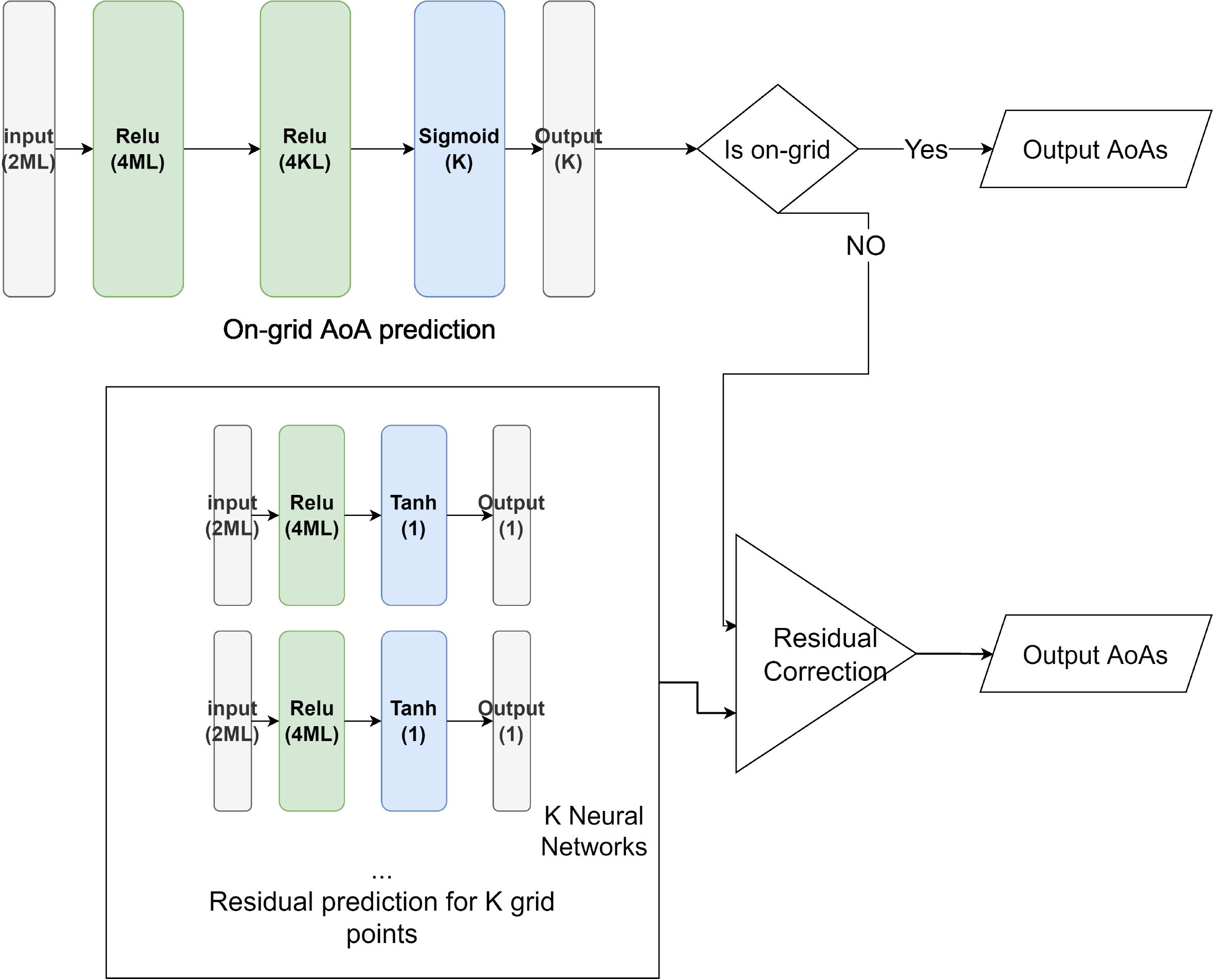}}
 	\caption{NN architecture for residual AoA prediction.}
 	\label{fig:delta_network}
 \end{figure}

 The NN architecture used to calculate the AoAs is shown in Fig. \ref{fig:delta_network}, where it consists of several NNs. First, the on-grid AoAs are predicted using the top NN in the diagram.  Stacked received signals over $L$ symbols is the input to this NN, and the residual AoAs are obtained at the output. The input feature is a complex vector, and we convert it to a real vector by stacking real and imaginary parts, which is fed to the network. At the output layer of this network \textit{sigmoid} activation is used, which corresponds to the probability of a certain AoA grid point being present.
 
 In the off-grid case, a collection of neural networks corresponding to each discrete grid point is used to predict the residual error in the AoA. These NNs are shown in the bottom part of the diagram. Each network has similar input as the on-grid AoA prediction network, while \textit{tanh} activation is used at the output layer. In this case, first the on-grid point is identified by the top network and the residual error is predicted using the corresponding bottom network. Finally, the AoAs are calculated by correcting the error.
 
%% file: results.tex
In this Section, we perform numerical simulations to evaluate the performance of the proposed algorithms. In all the simulations BS antenna is considered as a $16 \times 1$ uniform linear array (ULA) oriented in the \textit{x-axis}. The RIS is considered as a $16 \times 16$ uniform planar array (UPA) oriented in the \textit{y-axis}. The BS and RIS locations are fixed, and user locations are generated randomly. All the devices are assumed to be at the same elevation ($z=0$). The channel is generated based on a mmWave scattering model with random angles. The noise level has been set to -110 dBm and number of discrete AoA grid points ($K$) is set to $32$.

\begin{figure}[ht]
	\centerline{\includegraphics[width=\columnwidth]{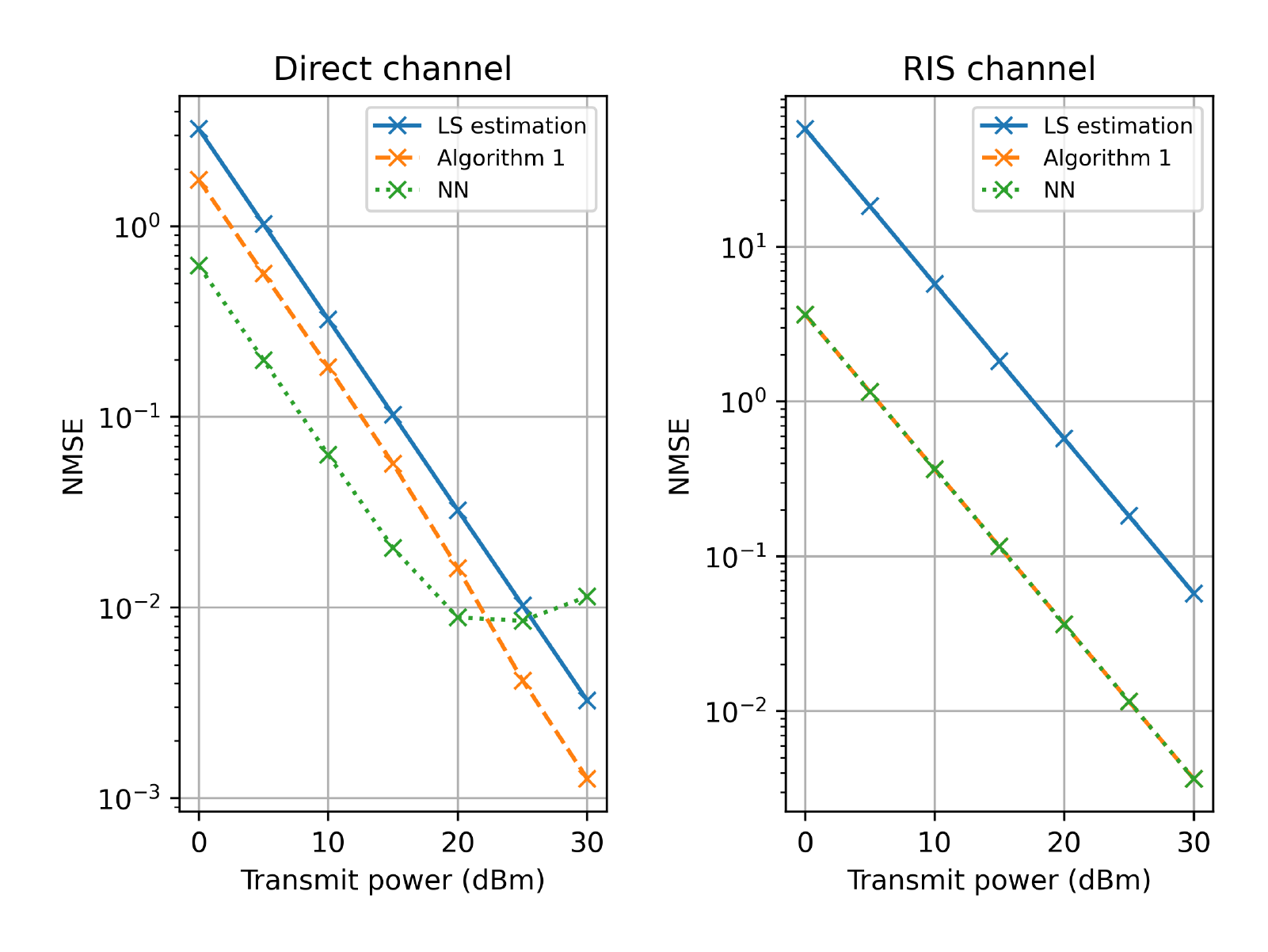}}
	\caption{Channel estimation with on-grid AoAs.}
	\label{fig:channel_estimation_ongrid}
\end{figure}

First, channel estimation with on-grid angles is simulated. Fig. \ref{fig:channel_estimation_ongrid} shows the plot of transmit power vs. normalized mean square error (NMSE). Results are generated using both Algorithm \ref{alg:1} and AoA prediction with the NN. Least squares (LS) estimation is also considered for comparison. Comparing the performance of direct channel estimation, we can see that Algorithm \ref{alg:1} outperforms LS at all power levels, while the performance gap also increases with increasing transmit power. The NN outperforms both LS and Algorithm \ref{alg:1}, but the performance get saturated at high transmit power. RIS channel is estimated by following the rest of the steps of Algorithm \ref{alg:1} in both cases, and it outperforms LS estimation.

\begin{figure}[ht]
	\centerline{\includegraphics[width=\columnwidth]{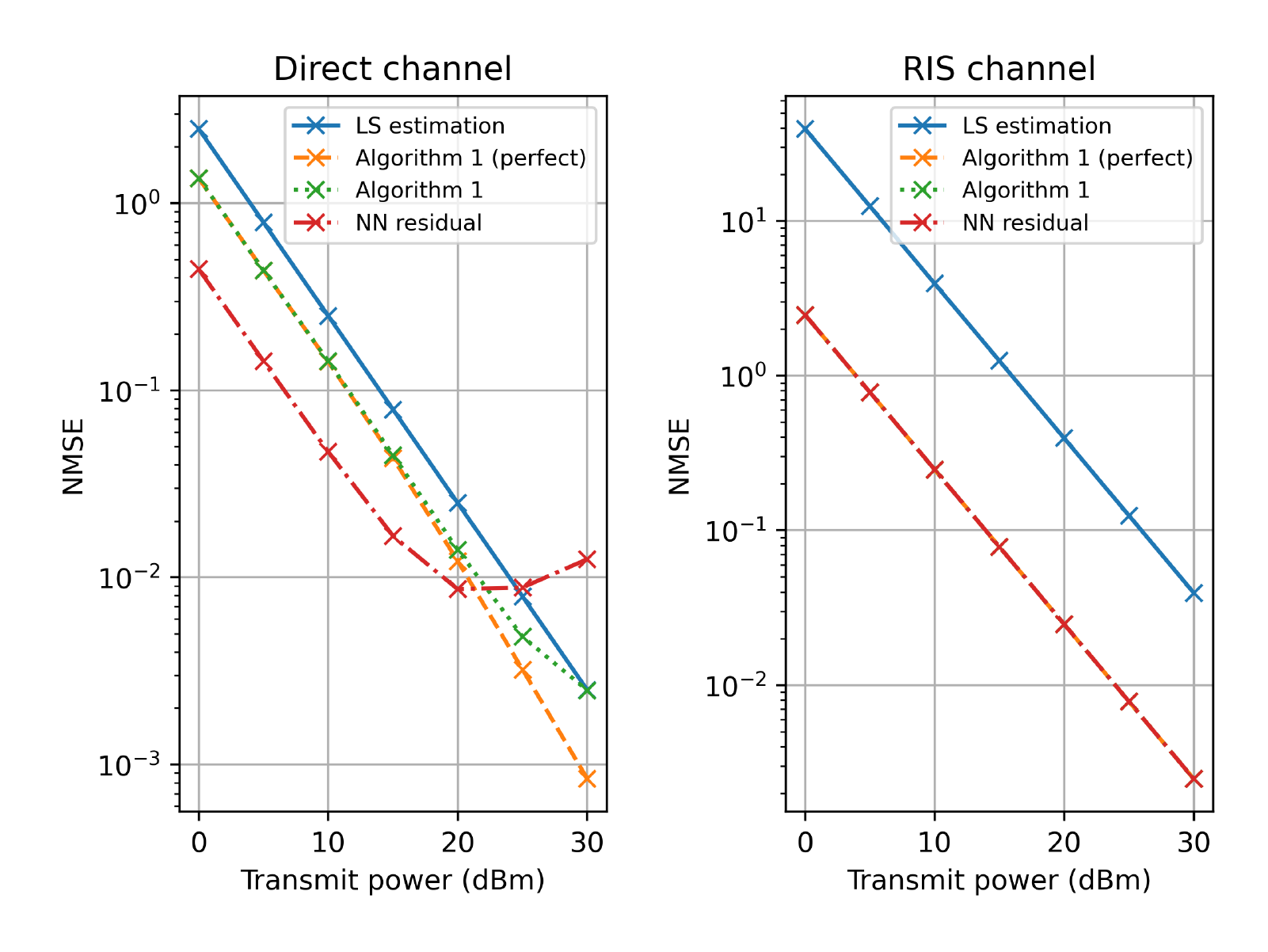}}
	\caption{Channel estimation with off-grid AoAs.}
	\label{fig:channel_estimation_offgrid}
\end{figure}

Next, off-grid AoAs are considered as shown in Fig. \ref{fig:channel_estimation_offgrid}. Performance of Algorithm \ref{alg:1} is evaluated under both perfect and imperfect AoA values, where in the perfect case, residual errors of AoAs are assumed to be perfectly known. Although AoAs are off-grid, we see a similar performance to the on-grid case. The NN based solution outperforms both LS and algorithms. However, in the direct channel, a saturation of performance is seen at high transmit power as observed earlier in the on-grid case. A similar effect is seen for Algorithm \ref{fig:channel_estimation_offgrid} with imperfect AoAs. This is related to the increasing power leakage due to grid imperfections. As a result the NN struggles to predict the correct AoA point. RIS channel estimation shows a similar performance as the on-grid case.

%% file: conclusion.tex
In this paper, we have considered the channel estimation of a mmWave network assisted by an RIS. We have formulated this as a sparse recovery problem by discretizing the AoAs. An algorithm was proposed to solve the on-grid case, along with NN based AoA prediction. In the off-grid case, another NN was used to predict the residual angles. Numerical simulations have shown that the proposed algorithm outperforms LS estimation. The NN network gives the best performance in both on-grid and off-grid cases, however, the performance gets saturated at high transmit power.  